\documentstyle[twocolumn,prb,aps,epsfig]{revtex}
\begin{document}

\twocolumn[\hsize\textwidth\columnwidth\hsize\csname @twocolumnfalse\endcsname

\title{Phase transition in the one-dimensional 
Kondo lattice model with attractive electron-electron interaction}

\author{Naokazu Shibata \cite{SHIBATA}, Manfred Sigrist and Elmar Heeb}

\address{Theoretische Physik, ETH-H\"onggerberg, 8093 Z\"urich,
  Switzerland}

\date{\today}

\maketitle

\begin{abstract}
  The one-dimensional Kondo lattice model with attractive interaction
  among the conduction electrons is analyzed in the case of
  half-filling.  It is shown that there are three distinct phases
  depending on the coupling constants of the model. Two phases have a
  spin and charge gap. While one shows a clear separation of the spin
  and charge excitation spectrum the other phase may be characterized
  as a band insulator type where both excitations are due to
  two-particle states.  The third phase is gapless in both channels
  and has quasi long-range order in the spin and charge density wave
  correlation. In this phase the spin and charge excitations have
  again a clearly separated spectrum.  For the analysis we discuss
  first two limiting cases. Then a density matrix renormalization
  group calculation on finite systems is applied to determine the
  phase diagram and the correlation functions in the gapped and
  gapless phase for general couplding constants.
\end{abstract}

\vskip2pc]

\narrowtext

\section{Introduction}

The Kondo lattice model (KLM) describes a many-body system of two
distinct types of degrees of freedom, itinerant electrons and
localized spins which are arranged on a regular lattice.  This model
can be considered as an extension of the single impurity Kondo model
where electrons interact with a single localized spin. The Hamiltonian
of the KLM consists of two parts, the kinetic energy of the itinerant
conduction electrons and the local exchange interaction between
electron spin $ {\bf S}_{\rm c} $ and localized spin $ {\bf S}_{\rm f}
$, both spin 1/2 degrees of freedom,

\begin{equation}
  {\cal H}_{\rm KLM} = -t \sum_{\langle i,j \rangle} \sum_s (
  c^{\dag}_{is} c_{js} + c^{\dag}_{js} c_{is}) + J \sum_i {\bf S}_{{\rm
      c}i} \cdot {\bf S}_{{\rm f}i}
\end{equation}
where the operator $ c_{is} $ ($ c^{\dag}_{is} $) annihilates
(creates) a conduction electron on site $ i $ with spin $ s $ ($=
\uparrow, \downarrow $) ($ S^{\mu}_{{\rm c}i} = (\hbar / 2)
\sum_{s,s'} c^{\dag}_{is} \sigma^{\mu}_{ss'} c_{is'} $) and the sum in
the first term runs over all nearest neighbor bonds $ \langle i,j
\rangle $.  Furthermore, $ t $ denotes the hopping matrix element and
$ J $ is the antiferromagnetic exchange coupling \cite{TSUNE1}.

In this model both the conduction electrons and the localized spins
are each uncorrelated. Correlation only appears through the exchange
interaction. It is important to realize that the understanding of the
single impurity Kondo problem does not simply extend to the lattice
case. Indeed the research during recent years has shown that
complicated correlation effects occur within the lattice model
yielding a variety of physical phenomena beyond the single impurity
picture.  Although the KLM is certainly of interest of its own as a
generic model of strongly correlated electrons, studies of this model
are also motivated by real materials such as the so-called heavy
fermion compounds or the Kondo insulators.  Both systems where
correlation effects between itinerant and localized electrons clearly
dominate the low-energy physics.

In this article we discuss the extension of the KLM obtained by
including the direct interaction among the conduction electrons. We
use the most simple form for the interaction, which acts only between
electrons on the same site,

\begin{equation}
  {\cal H}_{\rm int} = U \sum_i c^{\dag}_{i \uparrow} c_{i \uparrow}
  c^{\dag}_{i \downarrow} c_{i \downarrow}.
\end{equation}
We call $ {\cal H} = {\cal H}_{\rm KLM} + {\cal H}_{\rm int} $
Kondo-Hubbard model (KHM). This model was recently also considered in
connection with the ferromagnetic ground state away from half filling
\cite{YANAGI}.  In the following we will focus on the one-dimensional
(1D) system with a half-filled electron band.

The half-filled KLM is considered as a good starting point to
understand the Kondo insulators, a class of materials which show a
spin and a charge gap at low temperatures. In contrast to ordinary
band insulators the two gaps are different, indicating a separation of
the spin and charge degrees of freedom due to correlation effects. The
half-filled KLM shows indeed this type of properties. The properties
of the spins are dominated by short-ranged antiferromagnetic
correlations, i.e. this state can be considered as a spin liquid. A
particular feature of the 1D KLM is that the spin liquid state exists
for all finite values of $ J $. This could be shown numerically using
exact diagonalization \cite{TSUNE1} or, more recently, the density
matrix renormalization group technique \cite{YU} and the mapping to a
non-linear sigma model \cite{TSVELICK}. (Note that in higher
dimensions a transition between the spin liquid and an
antiferromagnetically ordered state is expected at a critical value of
$ J $. This has, however, not been established so far either by
analytical or numerical methods.) For weak coupling, $ J \ll t $, one
finds that the spin gap ($ \Delta_{\rm s} $) and charge gap ($
\Delta_{\rm c} $) depend in a very different way on $ J $,

\begin{equation} \begin{array}{l}
    \Delta_{\rm s} \propto \exp (- 1/ \alpha \rho J) \\ \\ 
    \Delta_{\rm c} \propto J \\ 
\end{array} \end{equation}
where $ \rho $ is the density of states at the Fermi level for the
free electrons. The spin gap energy gives an energy scale formally
related to the Kondo temperature $ T_K $ of a single localized
spin ($ \alpha = 1 $). For the lattice of localized spins $ \alpha $
is enhanced by a value of about 1.4 obtained from numerical
simulations \cite{SHIBATA1}. The similarity with the Kondo energy
scale indicates that the ground state of the half-filled KLM
corresponds to a singlet bound state like the Kondo singlet state.
However, correlation effects among the localized spins tend to
increase the binding energy through the formation of a collective
singlet state.  On the other hand, the charge gap has no counter part
in the single impurity case (where the system forms a Fermi liquid).
The charge gap proportional to $ J $ originates from the strong
antiferromagnetic correlation of the localized spins. Although short
ranged it provides a staggered background for the electron motion,
which yields the features of a doubled unit cell as found in a spin
density wave state \cite{REVIEW}.  Thus the energy scale is set by the
exchange coupling between the electrons and the localized spins.  Also
in the strong coupling limit ($ J \gg t $) we find a clear distinction
between the two excitations as we will consider below.

The spin liquid state of the KLM can be characterized as the formation
of a collective Kondo singlet involving the conduction electrons and
localized spins. The singlet formation is optimized if configurations
of conduction electrons with doubly occupied and empty sites are
suppressed. Both configurations ``remove'' the electron spin degree of
freedom so that the exchange term cannot be active.  Therefore strong
charge fluctuations of the conduction electrons tend to weaken the
spin liquid state. The inclusion of a repulsive interaction between
the conduction electrons as in Eq.(2) with $ U >0 $ leads to a
suppression of the charge fluctuations. It is a well-known fact that
the Hubbard model with repulsive interaction develops a charge gap at
half-filling for any finite $ U $ in one dimension, while the spin
excitation remain gapless. Consequently, we expect that the spin
liquid state is further stabilized by the repulsive interaction. This
is clearly seen in numerical calculations where, in particular, in the
limit of $ J \ll t $ the spin gap is enhanced through an increase of
the factor $ \alpha $ in the exponent Eq.(3) \cite{SHIBATA1}. While
the spin gap goes to zero for $ J \to 0 $ the charge gap remains
finite, if $ U >0 $.

We now ask what will happen if the interaction among the conduction
electrons is attractive.  There is a relation between the positive and
negative $ U $ Hubbard model due to particle-hole symmetry at half
filling. The change of sign for $ U $ leads effectively to an exchange
of the charge and spin degrees of freedom. Indeed the charge
excitations can be described as isospins completely analogous to the
spins, as we will see below \cite{AUERBACH}. For $ U < 0 $ the spin
sector of the conduction electrons has a gap while the charge
(isospin) excitations are gapless. Obviously this spin gap weakens the
spin liquid phase characterized by the formation of the singlets
between localized and electron spins. Therefore a competition arises
between the the spin and the charge fluctuations whose behavior is
determined by the relative strength of the coupling constants $ J $
and $ U $.

It is the goal of this paper to investigate this competition for the
case of $ U < 0 $. By the analysis of two limiting cases we will
demonstrate that the attractive interaction leads to a phase
transition where the character of the ground state and the excitations
change qualitatively. The state which is in competition with the spin
liquid phase may be characterized by the property that it has
quasi-long-range order in the spin and charge sectors. The dominant
correlations are that of an antiferromagnet for the localized spins
and of a charge density wave for the conduction electrons, i.e.  a
spin-charge density wave (SCDW). Additionally we find that within the
spin liquid a phase can occur where the spin and charge excitations
have the same energy scale and behave similar to a band insulator. In
the following we will analyze these properties first by characterizing
the states in the limit $ t \ll |U|, J $. In a second step we will
investigate the phase transition and the phase diagram, $ U $ versus $
J $, by means of numerical simulation using the density matrix
renormalization group method (DMRG).

\section{Phases in two limiting cases}

In this section we show that there are two distinct phases for the KHM
with attractive interaction. To this end it is helpful to consider two
limiting cases which allow a simple analysis of the model.  These
limits are $ t \ll J \ll - U $ and $ t \ll - U < J $.  At the very
beginning let us start with $ t=0 $, the atomic limit.  Then the
states can be represented most easily in a real space basis.  For $
U=0 $ the ground state $ | \Psi_{\rm s} \rangle $ is a product of
onsite singlets, i.e. on each site we find one conduction electron
which forms a spin singlet with the localized spin ($ |\Psi_{\rm s}
\rangle = (c^{\dag}_{i \uparrow} |\downarrow_i \rangle - c^{\dag}_{i
  \downarrow} | \uparrow_i \rangle) / \sqrt{2} $; $ |s_i \rangle $ is
the electron vacuum with the localized spin $ s_i $ on the site $ i
$). This state is {\it non-degenerate} \cite{TSUNE2}. The lowest spin
excitation $ 
|\Psi_{\rm t} \rangle $ ($ = c^{\dag}_{i \uparrow} | \uparrow_i
\rangle $) corresponds to the spin triplet configuration on one site.
Since one singlet had to be replaced by a triplet the excitation
energy is $ J $.  This state is highly degenerate ($ 3 N $) due to the
freedom of position and the three triplet orientations ($ N $ : number
of lattice sites). The lowest charge excitation $ |\Psi_{\rm c}
\rangle $ consists of a doubly-occupied and an empty site, which we
call doublon and holon, respectively ($ |\Psi_{\rm c} \rangle = |s_i
\rangle + c^{\dag}_{j \uparrow} c^{\dag}_{j \downarrow} |s_j \rangle
$).  The excitation energy is $ 3 J /2 $, because two singlets are
destroyed to generate this state.  The doublon and holon are fermionic
particles with a spin 1/2 degree of freedom, which combine to a spin
singlet for a pure charge excitation.  Then the degeneracy is $ N(N-1)
$. The combination of spin and charge configuration corresponds to the
triplet configuration of the doublon-holon spins with a degeneracy of
$ 3 N(N-1) $.

\begin{figure}
  \epsfxsize=80mm \epsffile{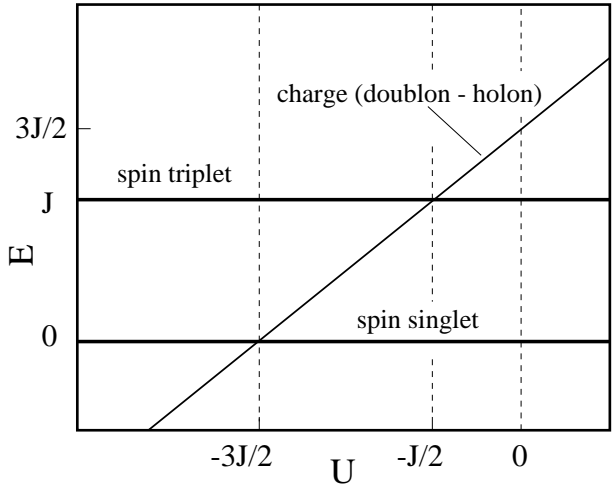}
\caption[*]{Level scheme of the states for $ t=0 $ as a function of the
  electron-electron interaction $ U $.}
\label{level}
\end{figure}
Turning on $ U $ now we can change the level scheme
(Fig.~\ref{level}). The relative position of the singlet and triplet
states, $ |\Psi_{\rm s} \rangle $ and $ | \Psi_{\rm t} \rangle $, is
unchanged.  However, the charge (doublon-holon) excitation is shifted
according to $ 3 J /2 + U $. For positive $ U $ no qualitative change
occurs in the level scheme. Negative $ U $, however, lead to a
rearrangement of the onsite energy levels for sufficiently large $ |U|
$. For $ U = - J/2 $ the charge excitation passes the triplet state
and for $ U = - 3 J /2 $ the singlet state. This indicates that the
attractive interaction between electrons yields a qualitative
modification of the system. Now let us discuss the situation where $ t
$ is finite and the above mentioned degeneracies are lifted.

\subsection{The spin liquid state: $ |U| \ll J $}

For small $ t $ we can now use perturbation theory to describe the
effect of hopping of the conduction electrons. A detailed discussion
of this type of perturbation can be found in Ref.\onlinecite{REVIEW}. 
For the limit $ |U| \ll J $ the singlet state $ |\Psi_{\rm s} \rangle $
remains the ground state \cite{TSUNE2}.  Its energy acquires a
correction in second 
order due to the polarization of the onsite singlet, i.e.  the
neighbor electron and localized spin are involved in the formation of
the singlet which becomes more extended.

The hopping term leads to the mobility of the spin triplet and the
doublon-holon excitations. The former behaves as a single
quasiparticle with an excitation energy

\begin{equation}
  E_{\rm t} (q) = J + \frac{4 t^2}{3J + 2U} - \frac{4t^2}{J+2U} (1 -
  \cos q)
\end{equation}
where $ q $ is the momentum (lattice constant $ a=1 $). The excitation
energy for the doublon-holon state has the form of two-particle
excitation with an energy depending on two momenta $ k $ and $ q $,

\begin{equation} \begin{array}{ll}
    E_{\rm c} (k,q) = & \displaystyle \frac{3J}{2} + U + \frac{8t^2}{3
      J +2U} - \frac{3 t^2}{J} \\ & \\ & \displaystyle + t \{
    \cos(k+q) + \cos(k) \} \\ & \\ & \displaystyle + \frac{t^2}{3 J +
      2 U} \{ \cos(2(k+q)) + \cos(2k) \}.  \\ &
\end{array} \end{equation}
Thus, the doublon-holon excitations form a continuum. The spin triplet
excitation may be understood also as a bound state of a doublon and
holon with their spins in the triplet configuration. In this sense the
triplet excitation has the character of an ``exciton'' within the gap
between the singlet ground state and doublon-holon continuum.

The discussion of the level scheme for $ t =0 $ suggests that there is
some change in the properties of the spin liquid phase as we turn $ U
$ from 0 towards negative values. There is a critical value $ U = - J
/ 2 $ where the doublon-holon state falls below the triplet excitation
(for $ t=0 $). If $ U $ is smaller than $ - J / 2 $ the triplet
excitation discussed above is absorbed into the continuum of the spin
triplet channel of the doublon-holon continuum. Therefore the lowest
spin and charge excitations have both two-particle character like a
particle-hole excitation. This state has still a finite gap and the
structure of the excitations is essentially similar to that of a band
insulator. The quasiparticles, the doublons and holons, are spin 1/2
fermions, composed of conduction electrons and localized spins.

At the second critical value of $ U $ ($ = - 3 J/2 $) the
holon-doublon state passes the singlet ground state. Here the
character of the ground state has to change so that this critical
value should belong to a phase transition. In the following we want to
discuss the properties of the new state.

The perturbative treatment of the different states given above does
not allow the evaluation of the transition points for finite $ t $.
Close to the transition points a more complicated perturbation for
degenerate states would be necessary. This leads, however, already to
a complicated many-body problem. Thus, we leave the discussion of the
phase boundary lines to the numerical part of our paper.

\subsection{The SCDW state: $ |U| \gg J $}

In the limit $ |U| \gg J $ the ground state is highly degenerate if we
assume $ t = 0 $. A complete basis of these degenerate states is given
by all real-space configurations of holons and doublons, with an equal
number of each (half filling). Both onsite singlets $ |\Psi_{\rm s}
\rangle $ or triplets $ | \Psi_{\rm t} \rangle $ are now excited
states which always should occur in pairs, due to the conservation of
the electron number, with energies $ E_{\rm ss} = -3 J/2 -U $, $
E_{\rm st} = -J/2 -U $ and $ E_{\rm tt} = J/2 -U $ for a
singlet-singlet, singlet-triplet and a triplet-triplet pair. All these
states are highly degenerate.

In order to lift the degeneracy of the ground state we include now the
hopping process, assuming a small, but finite $ t $. We can now
generate an effective Hamiltonian in the Hilbert-subspace of doublons
and holons only. At this point it is convenient to introduce the
notion of isospin $ {\bf I} $ (I=1/2) for the charge degrees of
freedom of the holon and doublon. We identify a holon with isospin up
and a doublon with isospin down. Like the ordinary spin also the
isospin transforms according to the SU(2)-symmetry group.
Additionally we introduce a phase convention by dividing the lattice
into two sublattices $ A $ and $ B $. Then the real space basis
functions should be multiplied by $ \prod_{i \in B} \exp (i \pi I_z)
$.

\begin{equation} \begin{array}{l}
    I^{+}_i = s_i c^{\dag}_{i \uparrow} c^{\dag}_{i \downarrow} \\ \\ 
    I^{-}_i = s_i c_{i \downarrow} c_{i \uparrow} \\ \\ \displaystyle
    I^z_i = \frac{1}{2}( c^{\dag}_{i \uparrow} c_{i \uparrow} +
    c^{\dag}_{i \downarrow} c_{i \downarrow} -1) \\ 
\end{array} \end{equation}
where $ s_i = +1 $ if $ i \in A $ and $ s_i = -1 $ if $ i \in B $.
With this notation the effective Hamiltonian which lifts the
degeneracy in lowest order perturbation theory has the form,

\begin{equation}
  {\cal H}_{\rm eff} = \sum_{\langle i,j \rangle} \left[ {\bf I}_i
    \cdot {\bf I}_j - \frac{1}{4} \right] \left[ a {\bf S}_{fi} \cdot
    {\bf S}_{fj} + b \right].
\end{equation}
where the two constants are

\begin{equation} \begin{array}{l} \displaystyle
    a = -\frac{ 2 t^2}{U+\frac{J}{2}} + \frac{t^2}{U-\frac{J}{2}} +
    \frac{t^2}{U+ \frac{3 J}{2}} , \\ \\ \displaystyle b =
    -\frac{9t^2/4}{U-\frac{J}{2}} - \frac{ 3 t^2/2}{U+ \frac{J}{2}} -
    \frac{t^2/4}{U+ \frac{3 J}{2}} \\ 
\end{array} \end{equation}
(derived in the Appendix).  In this formulation the SO(4)-symmetry of
both the spin and isospin is obvious (SO(4) = SU(2) $ \times $ SU(2)).
The spin degrees of freedom are only due to the localized spins while
the conduction electron spin is completely quenched in the effective
model. Thus the conduction electrons only appear as charge degrees of
freedom, i.e. isospins. It is easy to see that the coupling constant $
a $ is negative for $ U<-3J/2 $ and vanishes for $ J \to 0 $, because this
leads to a complete decoupling of the localized spins from the
conduction electrons. Note that the positive constant $ b $ determining the
spectrum of the conduction electrons remains finite for $ J = 0 $
so that the isospin degrees of freedom remain coupled.
The electron spins appear only in states including onsite singlet or
triplet components which are much higher in energy.

For this type of Hamiltonian the following two exact statements are
known.  (1) For a finite system we can show that the ground state is a
singlet in both the spin and the isospin channel and it is
non-degenerate. This can be proven using a generalization of
Marshall's theorem.  (2) In the thermodynamic limit the excitations in
both channels are either gapless or the ground state is degenerate
with spontaneously broken parity. This can be demonstrated using a
generalization of the Lieb-Schultz-Mattis theorem as given by Affleck
and Lieb \cite{LSM,AFFLIEB}. There is no need to reproduce the proofs
of these two statements here, since they are completely analogous to
those applied to the Heisenberg spin system.

The latter statement proves that in the limit $ |U| \gg J $ the phase
is different from the spin liquid state. While the spin liquid phase
has a unique ground state and a gap in all excitations, this phase has
either a degenerate ground state or gapless excitations.  In the case
of gapless excitations we expect that the system has quasi-long-range
order in both spin and isospin analogous to the 1D Heisenberg model.
The dominant correlation of both degrees of freedom is
``antiferromagnetic'', which corresponds to the usual charge density
wave correlation for the isospin part.  Due to the qualitative
difference of the two ground states, we conclude that there must be a
phase transition between the two limits, depending on $ U $ and $ J $.
In the following we will use numerical methods to demonstrate that the
state for $ U \ll - J $ has actually gapless excitations.

\section{The phase diagram by DMRG}

In order to determine the transition line in the phase diagram, $ U $
versus $ J $, and the character of the different phases, we calculate
the excitation energies and the correlation functions of the ground
state in the KHM numerically. For this purpose it is necessary to
treat long enough systems such that the transition between different
phases can be observed reliably.  We use the density matrix
renormalization group (DMRG) method which allows us to study long
chains by iteratively enlarging the system size and obtain ground
state wave functions with only small systematic errors. \cite{WHITE}

We calculate the elementary excitations for various $U$ to identify
the three phases discussed in Sec.~II.  The DMRG scheme is designed to
obtain a very good approximation of the ground state of a model. For
the evaluation of excited states and their energies we consider the
model in those Hilbert-subspaces which contain these states as lowest
energy states.  The real ground state is found for the subspace with
$S_{\rm tot}^z=0$ and $N_{\rm c}=L$ ($ N_{\rm c} $: number of
conduction electrons).  The spin excited state is obtained for $
S_{\rm tot}^z =1 $ and $N_{\rm c}=L$ and the charge excited state for
$S_{\rm tot}^z=0$ and $ N_{\rm c} = L \pm 2 $.  In these sectors the
lowest energy state is calculated for various system sizes up to
$L=48$ with the finite system algorithm using open boundary
conditions. The extrapolation to the bulk limit is obtained from the
scaling laws. For the gapped spin liquid phase the form
$\Delta(L)=\Delta(\infty)+\beta L^{-2}+O(L^{-4})$ is assumed, since
the lowest excited state generally corresponds to the bottom of an
excitation spectrum which can be expanded in terms of the square of
the momentum, $k^2$.  On the other hand, for the region which is
supposed to possess the gapless phase the bulk limit value of the
excitation energies is estimated simply using $1/L$ scaling and
$1/\log{L}$ scaling.

\begin{figure}
  \epsfxsize=85mm \epsffile{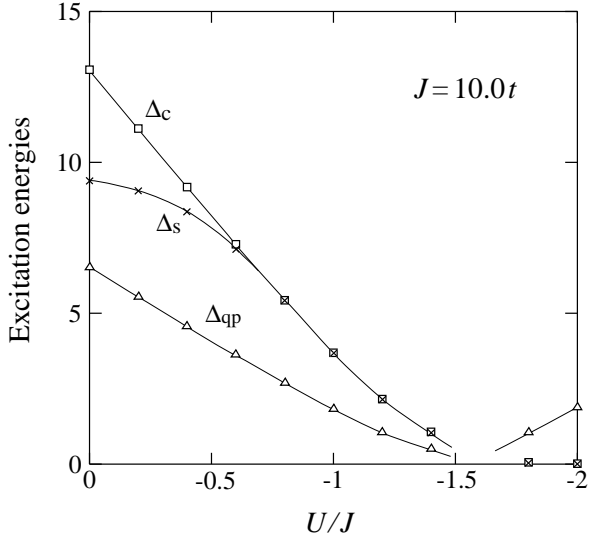}
\caption[*]{Excitation energies of the half-filled Kondo lattice model
  with attractive interactions.  Typical truncation errors in the DMRG
  calculations are $1 \times 10^{-6}$ for $U/J=-1.4$.  Gap energies,
  exchange constant $J$, and Coulomb interaction $U$ are in units of
  $t$.  }
\label{exci10}
\end{figure}
\begin{figure}
  \epsfxsize=85mm \epsffile{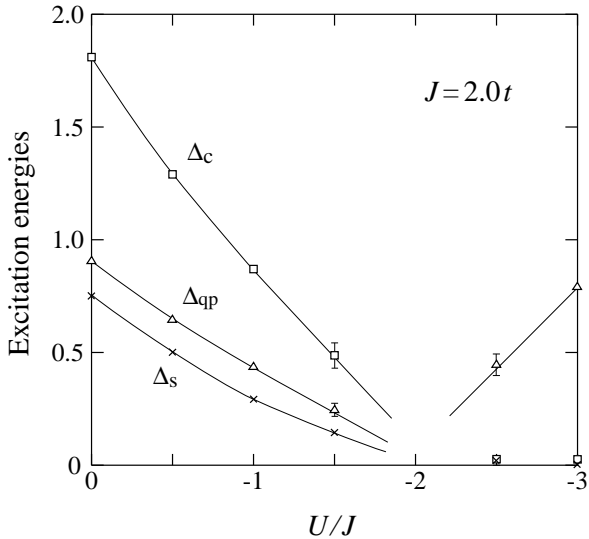}
\caption[*]{
  Excitation energies of the half-filled Kondo lattice model with
  attractive interactions.  Typical truncation errors in the DMRG
  calculations are $5 \times 10^{-5}$ for $U/J=-1.5$.  Errorbars are
  estimated from $1/L^2$ and $1/\log{L}$ scalings.  Gap energies,
  exchange constant $J$, and Coulomb interaction $U$ are in units of
  $t$.  }
\label{exci02}
\end{figure}
First let us discuss the results for the case of strong exchange
coupling $ J=10.0t$. The data for the excitation gaps are shown in
Fig.~\ref{exci10}. It is not difficult to identify the three phases as
we scan $ U $ from 0 to $ -2 J $: the ordinary spin liquid phase
$\Delta_{\rm c}>\Delta_{\rm s}>0$ for $U/J>-0.6$, the spin liquid
phase with identical spin and charge excitations $\Delta_{\rm
  c}=\Delta_{\rm s}>0$ for $-0.8>U/J>-1.4$, and the gapless phase
(SCDW) $\Delta_{\rm c}=\Delta_{\rm s}=0$ for $U/J<-1.8$. In the
intermediate regime located between $U/J=-0.8$ and $-1.2$ the
difference of $\Delta_{\rm c}$ and $\Delta_{\rm s}$ vanishes in the
bulk limit within the accuracy of the present calculation. Therefore
this confirms the existence of the intermediate phase where both spin
and charge excitations have the same energy properties and are
essentially of the two particle type as anticipated in the previous
section.

In addition the single quasiparticle excitation gap $\Delta_{\rm qp}$
is shown in Fig.~\ref{exci10}.  The quasiparticles originate from
doublons or holons and have therefore fermionic character.  The
$\Delta_{\rm qp}$ is obtained from the difference of the lowest energy
in the Hilbert spaces with $S_{\rm tot}^z=0$ with $N_{\rm c}=L$ and
$S_{\rm tot}^z=\pm 1/2$ with $N_{\rm c}=L\pm 1$. The $\Delta_{\rm qp}$
is half of the $\Delta_{\rm c}$ for $U/J>-1.4$. The reason is that the
effective interaction between the quasiparticles is repulsive in that
region. On the other hand, for $U/J<-1.8$ the $\Delta_{\rm qp}$
increases with growing $-U/J$ while $\Delta_{\rm c}$ is zero (within
our accuracy). Hence, we may interpret this as a switching from a
repulsive to an attractive quasiparticle interaction, indicated by the
minimum of the $\Delta_{\rm qp}$ around $U/J=-1.6$. Since the
character of the excitation changes here we expect that this switching
coincides with the phase transition to the gapless phase. Obviously,
in the thermodynamic limit all gaps should disappear at the transition
point.  Despite careful scaling analysis it is difficult to determine
the exact position of the transition from the present DMRG study
because close to the transition point the truncation error in the
density matrix becomes large and the convergence of the RG iteration
is rather slow.

Next we turn to a weaker exchange coupling.  The results for $J=2.0t$
are shown in Fig.~\ref{exci02}.  In contrast to the strong coupling
case we cannot find any indication of the intermediate regime with $
\Delta_{\rm c} = \Delta_{\rm s} > 0 $. This means that this regime is
either absent or confined to a very small region close to the
transition point which we locate at around $U/J \approx -2.0t$. For
$U/J < -2.0t$ the two gaps essentially coincide and are practically
zero. Corresponding to this change the quasiparticle gap shows a
minimum around the critical value of $ U $.

\begin{figure}
  \epsfxsize=85mm \epsffile{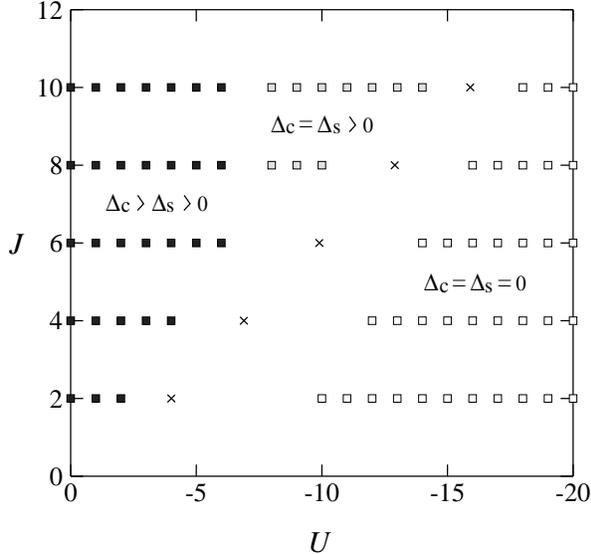}
\caption[*]{
  Ground state phase diagram of the half-filled Kondo lattice model
  with attractive interactions.  The cross represents the point where
  the quasiparticle gap becomes minimum.  Exchange constant $J$ and
  Coulomb interaction $U$ are in units of $t$.  }
\label{JU-phase}
\end{figure}

Similar calculations are carried out for $J/t=4,6,8$ and the phase
diagram is obtained.  In Fig.~\ref{JU-phase} we show the three phases
by different dots which are numerically determined. The crosses in
Fig.~\ref{JU-phase} represent estimated minimum points in the
quasiparticle gap where we expect the phase transition to be located.
There are technical limitations on the details of the phase diagram,
which do not allow us to determine clearly the extension of the
intermediate phase ($ \Delta_{\rm c} = \Delta_{\rm s} > 0 $).

A definite advantage of the DMRG scheme is that we can get very good
approximations of the ground state wave functions for rather large
systems. This allows the study of various correlation functions at
least sufficiently far from the transition line of the phase diagram.
Here we have analyzed the spin-spin and charge-charge correlation
functions which show characteristic features of the phases. The
correlation functions can be observed through the perturbing effect of
the boundaries.  This induces an oscillating disturbance into the wave
function leading to a charge or spin density modulation analogous to
the Friedel oscillations around an impurity.  For example if
$\Delta_{\rm c}$ is finite then the density-density correlation has an
exponential form and the charge density Friedel oscillations induced
by an impurity potential shows the same exponential decay. The length
scale is related to the ratio between the charge velocity and the gap,
$v_{\rm c}/\Delta_{\rm c}$.  On the other hand, if $\Delta_{\rm c}$ is
zero, then a power law decay of the correlation function, similar to
that of a Tomonaga-Luttinger liquid, is expected. This power law decay
occurs naturally in the Friedel oscillations of both spin and charge.

\begin{figure}
  \epsfxsize=85mm \epsffile{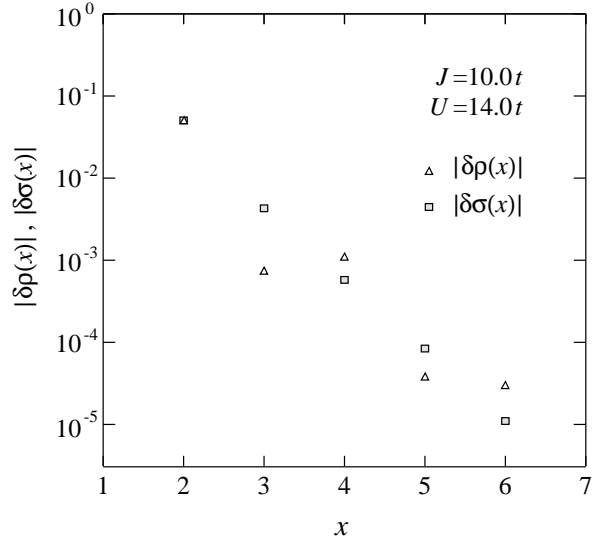}
\caption[*]{
  Charge density oscillations induced by the open boundary conditions,
  and spin density oscillations introduced by applying a local
  magnetic field.  The system size is 30 sites, $J=10t$, and $U=-14t$.
  The strength of the local magnetic field $h$ is $0.2t$.  }
\label{exp-dec}
\end{figure}

In Fig.~\ref{exp-dec} we show the charge and spin density Friedel
oscillations, $\delta\rho(x)$ and $\delta\sigma(x)$, in the
intermediate phase with $J/t=10$ and $U/J=1.4$. The charge density
oscillations are naturally induced by the open boundary conditions,
while for generating the spin density oscillations we have to apply a
local magnetic field, $H_{local}=2h(S_1^z-s_1^z-S_L^z+s_L^z)$,
coupling to the spins at both ends of the finite system.  Obviously
both the charge and spin density oscillations decay exponentially and
our analysis shows that their correlation length is essentially the
same.  This means that not only the excitation gaps but also the
velocities of the excitations are identical as we expect for the
excitations of the particle-hole type.  This clearly indicates that
the picture of this phase as a kind of band insulator is appropriate.

\begin{figure}
  \epsfxsize=85mm \epsffile{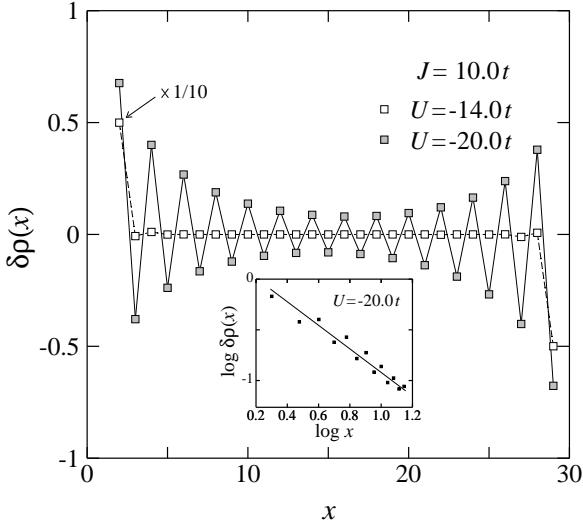}
\caption[*]{
  Charge density oscillations of the half-filled Kondo lattice model
  with attractive interactions.  The system size is 30 sites and
  $J=10t$.  The solid line and the broken line correspond to $U=-20t$
  and $-14t$, respectively.  The inset shows log-log plot of the
  oscillations for $U=-20t$.  }
\label{chrge-osc}
\end{figure}

\begin{figure}
  \epsfxsize=85mm \epsffile{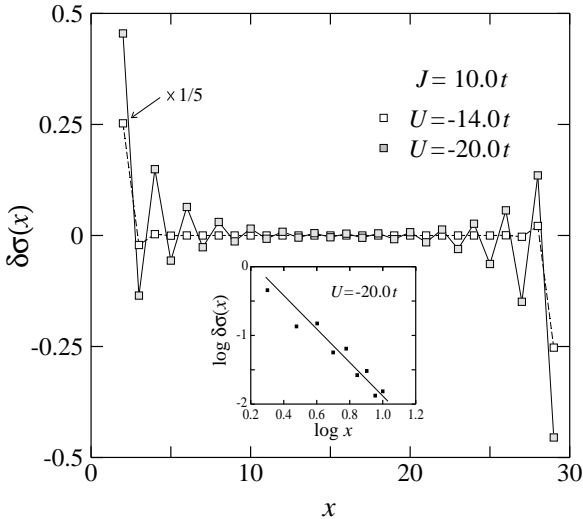}
\caption[*]{
  Spin density oscillations of the half-filled Kondo lattice model
  with attractive interactions.  The system size is 30 sites and
  $J=10t$.  The solid line and the broken line correspond to $U=-20t$
  and $-14t$, respectively.  The strength of the local magnetic field
  $h$ is $0.2t$.  The inset shows log-log plot of the oscillations for
  $U=-20t$.  }
\label{spin-osc}
\end{figure}

The charge and spin density Friedel oscillations in the gapless phase
are shown in Fig.~\ref{chrge-osc} and \ref{spin-osc}, respectively, by
the solid line.  In contrast to the exponential decay in the gapped
phase, represented by the broken line, a power law decay is observed
here, $ \sim (-1)^r / r^{\eta_{\rm c,s}} $.  Considering the insets of
Fig.~\ref{chrge-osc} and \ref{spin-osc} we find that the powers $
\eta_{\rm c,s} $ of decay are different for spin and charge (for the
charge $ \eta_{\rm c} \approx 1.15 $ and for the spin $ \eta_{\rm s}
\approx 2.4 $). This demonstrates the separation of spin and charge
excitations as anticipated from the effective Hamiltionian in Eq.(7).
In addition the staggering correlations both for spin and charge
densities is obviously dominating. This is in agreement with the
characterization of this phase as a spin and charge density wave with
quasi long-range order, as the effective Hamiltonian in Eq.(7)
suggests.  From this result it is clear that among the two
possibilities the Lieb-Schultz-Mattis theorem proposes for the phase $
-U \gg J $ the gapless state is the correct choice.

\section{Conclusion}

We have seen that the 1D half-filled Kondo-Hubbard lattice with
attractive electron-electron interaction has three phases, two gapped
phases (spin liquid and band insultator-like) with short range
correlations and one gapless phase with quasi long-range ordered spin
and charge density waves (SCDW). Note that for positive $ U $ only the
spin liquid exists. An indication of the difference between positive
and negative $ U $ can be found in the small-$ J $ limit. One may be
tempted to lift the large spin degeneracy found for $ J =0 $ by
perturbation theory, i.e. introducing the RKKY-interaction among the
localized spins. In lowest order one separates the system into free
conduction electrons and interacting localized spins.  A simple
calculation shows, however, that this perturbation concept fails for
the 1D system, because the effective interaction between the spins
diverges for the wave vector $ q = 2 k_F = \pi $ for all $ U \geq 0 $.
This indicates that there is no separation between these degrees of
freedoms and numerical calculations suggest that for any finite $ J $
the ground state has spin liquid properties. In contrast, for negative
$ U $ the perturbation converges for all wave vectors and one can
derive a sensible RKKY-model \cite{schoeller}.  Under this condition
the localized spins and the conduction electrons can separately
exhibit gapless phases. Note, however, that the electron spin
excitation has a gap for negative $ U $ and only the charge part is
gapless with a tendency towards an enhancement of charge density wave
correlations.  The effective Hamiltonian (7) describes this kind of
system in the extreme limit ($ |U| \gg J $).

An interesting change occurs also for the nature of the excitations.
If we scan from $ U = 0 $ to $ U \gg J $ for fixed values of $ t $ and
$ J $ then we find close to the transition ($ U \approx - 3 J/2 $) a
region where the lowest spin and charge excitations are due to
doublon-holon excitation. Therefore, the two excitations have the same
spectrum as a band insulator.  We would like to point out, however,
that the fermions, the holons and doublons, building the excitations
are quasiparticles which are composed of the itinerant electronic and
the localized spin degree of freedom. For finite $ t $ they are real
dressed quasiparticles involving complex correlation effects.  Thus it
is not an entirely trivial feature of this strongly correlated system
to mimic the properties of a simple band insulator.  On the other
hand, in the spin liquid phase for $ |U| \ll J $ the spin and charge
excitation have different nature.  The spin triplet excitation has
excitonic character, i.e. it is a bound state of a holon-doublon
excitation.

Also in the gapless phase the spin and charge excitations are
separated as is seen in the difference of the correlation functions.
However, the origin of the spin-charge separation is different. The
conduction electron and localized spins provide nearly independent
degrees of freedom in close connection to the RKKY-model mentioned
above.

The KHM with attractive interaction is an example of a strongly
correlated electron system which possesses a number of phase of rather
different character. These phases depend only on the relation of
different coupling constants. Several of the mentioned features can be
transfered to higher dimensional systems. The gapless ground state has
very likely long-range order in both the charge and the spin density
wave correlation. However, it has to be noticed that the spin liquid
phase for $ U \geq 0 $ might have antiferromagnetic long-range order,
if $ J $ is sufficiently small.  This has not been proven so far.
However, this long-range ordered state has a charge gap (gap of a spin
density wave state) and the spin excitations are due to both the
electron and the localized spin. This is in contrast to the SCDW-state
we discussed above. Thus in higher dimensions we may expect to see a
richer phase diagram.

\acknowledgements

We would like to thank to H. Tsunetsugu, K. Ueda and H. Schoeller for
helpful discussions. This work is financially supported by the Swiss
Nationalfonds. M.S. is grateful for the PROFIL-fellowship by the Swiss
Nationalfonds.

\begin{appendix}
\section{Effective Hamiltonian in the limit $|U|\gg J\gg t $ }

The low-energy Hilbert subspace $ \Lambda $ in the limit of large
attractive interaction consists of all real-space configurations of
holons and doublons which appear in equal number at half-filling. All
other states containing onsite singlets or triplets are higher in
energy by a multiple of order $ |U| $.  For convenience we introduced
the notion of isospin $ \tilde{{\bf I}} $ as an additional local SU(2)
degree of freedom besides the localized spins. The doublon and the
holon are mapped to isospin up and down, respectively (Eq.(6)). Thus
each site has four basis states: $ | \tilde{I}^z, S^z \rangle = \{ |+
, \uparrow \rangle , |+ , \downarrow \rangle, |- , \uparrow \rangle ,
|- , \downarrow \rangle \} $. For vanishing hopping $ t $ the whole
subspace $ \Lambda $ is disconnected leading to complete degeneracy.
We would now like to generate the effective Hamiltonian which lifts
this degeneracy and determines the low-energy physics. We use second
order perturbation theory in the hopping where the higher energy
states outside of $ \Lambda $ appear as intermediate states. In this
way the effective Hamiltonian will only have nearest neighbor coupling
and be of the generic form,

\begin{equation} \begin{array}{ll}
    {\cal H}_{\rm eff} = \sum_i & \displaystyle \{ a \tilde{I}^z_i
    \tilde{I}^z_{i+1} + \frac{b}{2} ( \tilde{I}^+_i \tilde{I}^-_{i+1}
    + \tilde{I}^-_i \tilde{I}^+_{i+1} ) +c \} \\ & \\ & \quad
    \displaystyle \times \{ a' S^z_i S^z_{i+1} + \frac{b'}{2} ( S^+_i
    S^-_{i+1} + S^-_i S^+_{i+1} )+ c' \}.
\end{array} \end{equation}
We consider now the coupling on a single bond, say between site 1 and
2, where we represent the local states by $ | \tilde{I}^z_1,
\tilde{I}^z_2 ; S^z_1, S^z_2 \rangle $. There are three different
cases to take into account:

\vskip 0.3 cm

\noindent
1) $ | \tilde{I}, \tilde{I} ; S ,\pm S \rangle $: If the $ z
$-component of the isospins on a bond are the same, then the hopping
is ineffective and this state is not connected to any other via this
bond.

\vskip 0.3 cm

\noindent
2) $ | \tilde{I},- \tilde{I} ; S, S \rangle $: We consider the example
$ | + ,-; \uparrow \uparrow \rangle $ which leads to the matrix
elements,

\begin{equation} \begin{array}{ll} 
    m_1 & \displaystyle 
     = \langle + , - ; \uparrow \uparrow | {\cal H}_{\rm eff} | +
    , - ; \uparrow \uparrow \rangle =
    (-\frac{a}{4}+c)(\frac{a'}{4}+c')
\\ & \\ & \displaystyle = \langle + , - ;
    \uparrow \uparrow | {\cal H}_{\rm eff} | - , + ; \uparrow \uparrow
    \rangle = \frac{b}{2}(\frac{a'}{4}+c')
\\ & \\ & \displaystyle = \frac{t^2}{U-\frac{J}{2}} +
    \frac{t^2}{U+\frac{J}{2}} < 0 \\ &
\end{array} \end{equation}
where the intermediate states consist of two triplets ($ E = -U +
J/2 $) or of one singlet and one triplet ($ E= -U - J/2 $).

\vskip 0.3 cm

\noindent
3) $ | \tilde{I},- \tilde{I} ; S, -S \rangle $: It is sufficient to
consider the example $ | + ,-; \uparrow \downarrow \rangle $,

\begin{equation} \begin{array}{ll} 
    m_2 & \displaystyle 
= \langle + , - ; \uparrow \downarrow | {\cal H}_{\rm eff} |
    + , - ; \uparrow \downarrow \rangle = (- \frac{a}{4} +c)(-
    \frac{a'}{4} + c') 
\\ & \\ & \displaystyle = \langle + , - ;
    \uparrow \downarrow | {\cal H}_{\rm eff} | - , + ; \uparrow
    \downarrow \rangle = \frac{b}{2} (- \frac{a'}{4} + c')
\\ & \\ & \displaystyle = \frac{1}{4} \left[
      \frac{5t^2}{U-\frac{J}{2}} + \frac{2t^2}{U+\frac{J}{2}} +
    \frac{t^2}{U+ \frac{3 J}{2}} \right] < 0 \\ &
\end{array} \end{equation}
and

\begin{equation} \begin{array}{ll} 
    m_3 & \displaystyle 
= \langle + , - ; \uparrow \downarrow | {\cal H}_{\rm eff} |
    + , - ; \downarrow \uparrow \rangle = (- \frac{a}{4}+c) \frac{b'}{2}
\\ & \\ & \displaystyle = \langle + , - ;
    \uparrow \downarrow | {\cal H}_{\rm eff} | - , + ; \downarrow
    \uparrow \rangle = \frac{b}{2} \frac{b'}{2} 
\\ & \\ & \displaystyle = - \frac{1}{4} \left[
      \frac{t^2}{U-\frac{J}{2}} - \frac{2t^2}{U+\frac{J}{2}} +
      \frac{t^2}{U+ \frac{3 J}{2}} \right] > 0. \\ &
\end{array} \end{equation}
Note that $ m_1 - m_2 = m_3 $. Now we can determine the coefficients
in Eq.(7). Fixing $ a = 1 $ we find $ b = -1 $, $ c = -1/4 $, $ a' =
b' $ and

\begin{equation} \begin{array}{l}
    a'= 4 (m_2 - m_1) < 0 \\ \\ c'= -(m_1 + m_2) >  0. \\ 
\end{array} \end{equation}
Next, we rotate the isospin on every second site by $ \pi $ around the
$ z $ -axis, $ \tilde{I}^{x,y} \to I^{x,y} = - \tilde{I}^{x,y} $ and $
\tilde{I}^z \to I^z = \tilde{I}^z $ by applying the phase factor $
\exp (i \pi \tilde{I}^z_i) $. With this phase convention on the basis
states $ b \to - b $ and the effective Hamiltonian reaches its
apparently SU(2) rotational symmetric form in both spin and isospin
space as given in Eq.(7).
\end{appendix}

\end{document}